# Study of residual stress in reactively sputtered epitaxial Si-doped GaN films

Mohammad Monish[*] and S. S. Major[†]

*Department of Physics, Indian Institute of Technology Bombay, Mumbai 400076, India*

## *Abstract*

Si-doped GaN films were grown on *c*-sapphire by rf magnetron reactive co-sputtering of GaAs and Si at various partial pressures of $N_2$ in Ar-$N_2$ growth atmosphere and their epitaxial character was ascertained by phi-scans. Energy dispersive x-ray spectroscopy revealed ~2 at.% Si in all the films, but the N/Ga ratio decreased substantially as $N_2$ percentage was reduced from 100% to 10%. High resolution x-ray diffraction revealed the dominant presence of edge dislocations (~$10^{12}$ cm$^{-2}$) in the films grown at 30% - 100% $N_2$, which decreased to ~$5 \times 10^{11}$ cm$^{-2}$ at lower $N_2$ percentages, at which, the density of screw dislocations was found to increase and attained values comparable to that of edge dislocations. The lattice parameters (*a* and *c*) were obtained independently to determine the in-plane and out-of-plane components of strain in films, which were analyzed to deduce the hydrostatic and biaxial strain contributions. The film grown at 100% $N_2$ displayed large micro-strain and hydrostatic strain due to excess/interstitial nitrogen, both of which decreased with the initial reduction of $N_2$ percentage, but increased again below 30% $N_2$ due to Ar incorporation. The films grown above 75% $N_2$ displayed compressive biaxial stress, which is attributed to possible nitrogen incorporation into grain boundaries and tensile side of edge dislocations. The reversal of biaxial stress to tensile character in films grown below 75% $N_2$ is explained by the prevalence of in-plane tensile stress generated during coalescence. The tensile stress decreased in films grown below 30% $N_2$, which is ascribed to the Ar incorporation and voided morphology. The presence of Si in the films does not have a significant influence on strain behaviour and appears to be masked by the dominant growth-related intrinsic effects.

Keywords: GaN, Si-doped GaN, reactive sputtering, epitaxy, residual stress

---

[*] monish.iitb@gmail.com
[†]Author to whom any correspondence should be addressed: syed@iitb.ac.in

## 1. Introduction

GaN and related III-V alloys have received exceptional attention as futuristic semiconductor materials, owing to the wide range of their current and potential applications, extending from short wavelength optoelectronics and white light sources to high power and high frequency transistors [1-4]. The ever-increasing scope of GaN calls for exploring non-conventional approaches of material growth and controlled in-situ doping, as well as the improved understanding of its structural properties, particularly those obtained by alternative methods. GaN films have been grown by metal-organic chemical vapor deposition (MOCVD/MOVPE) and molecular beam epitaxy (MBE) mostly on sapphire substrate, despite the substantial mismatch between lattice constants and coefficients of thermal expansion [5]. The heteroepitaxial GaN films have often revealed high density of threading dislocations in the range of $10^8$ - $10^{10}$ cm$^{-2}$ [6-8], but unlike most other compound semiconductors, have been successfully utilized to fabricate high quality optoelectronic devices [9, 10]. However, due to limitations of high temperature operation in MOCVD and the scalability and high cost of MBE, alternative methods, such as sputtering have become attractive and continue to be explored for the growth of GaN [11-15]. In recent years, sputtering has demonstrated considerable promise for the growth of epitaxial GaN films, albeit with threading dislocations in the range of $10^9$-$10^{12}$ cm$^{-2}$ [16-18] and has been employed for obtaining undoped [16, 18, 19] and doped (with Si, Ge and Mg) [17, 20-22] GaN films, as well as fabrication of GaN based LEDs [23].

Silicon (Si) is a commonly used n-type dopant in GaN, yielding carrier concentrations as high as ~$2 \times 10^{20}$ cm$^{-3}$ in the films grown by different techniques and is used in various device structures [20, 24-26]. In an earlier work from our group, Si-doped GaN films have been grown by reactive co-sputtering of GaAs and Si [21], displaying resistivity in the range of $10^{-3}$ - 1 Ω-cm, depending on the Si area coverage of GaAs target and N$_2$ percentage in Ar-N$_2$ growth atmosphere. More recently, sputtered Si-doped GaN films [27], containing ~2 at.% Si have been grown by this approach, which revealed concurrent increase of carrier concentration and mobility with reduction of N$_2$ percentage in growth atmosphere from 100% to 10%, and displayed high conductivity, with carrier concentration ≳$10^{20}$ cm$^{-3}$ and mobility ≳20 cm$^2$V$^{-1}$s$^{-1}$ in films grown at ~10% N$_2$. As GaN based optoelectronic and electronic devices are mostly fabricated on sapphire or Si substrates, the presence of substantial strain in doped GaN films often influences their electrical and optical behaviour and is a potential cause of degradation in device performance [28]. The covalent radius of Si (1.11 Å) is smaller than that of Ga (1.26 Å), which is normally expected to cause a decrease in lattice parameters. It has however been shown [29, 30] by first-principles calculations that the net strain due to the substitution of Si in GaN lattice is negligibly small, since the effects due to size and deformation potentials are nearly similar in magnitudes but opposite in sign. Strain measurements of Si-doped GaN films on sapphire by several workers [29, 31-38] have also found marginal effects due to Si incorporation, but on the other hand, have widely reported the relaxation of in-plane (biaxial) compressive stress, or its change to tensile character, as will be summarized below. It may be mentioned here that measurement of strain in Si-doped GaN films grown by alternative approaches like sputtering, has not been reported.

Early studies of strain in MOCVD/MOVPE grown Si-doped GaN films on sapphire by out-of-plane XRD measurements revealed the presence of compressive biaxial stress, which relaxed with the incorporation of Si [31-34]. Cremades *et al.* [35] used XRD measurements of symmetric and asymmetric reflections to independently deduce the *a* and *c* parameters of MOCVD grown Si-doped GaN with different doping concentrations and found that the incorporation of Si in GaN not only led to stress relaxation, but also induced in-plane tensile stress in the films with higher carrier concentrations ($n > 10^{18}\,cm^{-3}$). They also observed a shift towards higher lattice constants and ascribed it to tensile hydrostatical pressure due to point defects [35]. Romano *et al.* [29] have also investigated the doping effect of Si on the microstructure and strain behavior in MOCVD grown GaN films by obtaining the in-plane and out-of-plane components of strain and have observed the reversal of compressive biaxial strain to tensile, with increase of Si concentration. They have ascribed the strain reversal to crystallite coalescence effects, in accordance with the Nix and Clemens model for polycrystalline films [39], while ruling out any effect due Si incorporation. The stress studies in Si and Ge doped GaN films have also been shown [36] to result in a relatively larger increase of in-plane tensile strain with Si concentration. However, Xie *et al.* [37] have argued that the tensile strain in Si-doped GaN films depends on the free carrier concentration. In another study on Si-doped AlGaN films, Forghani *et al.* [38] have proposed that the tensile strain in Si-doped (Al)GaN epitaxial films may be attributed to the incorporation of Si on the compressive side of edge dislocations. The above studies show that Si doping in GaN films on sapphire usually induces in-plane tensile stress, although the reasons for the same are not evident. It may however be noted that growth-related intrinsic stresses, such as, those generated during coalescence or due to point defects and microstructure related effects have not been considered in most of these studies, but may have a significant role in determining the residual stress in Si-doped GaN films. It may be relevant to note that the investigations of undoped GaN films grown on sapphire by different methods by Kisielowski *et al.* [40] have shown that both biaxial and hydrostatic strains may be concurrently present in these films. Joachim *et al.* [41] have also shown the presence of hydrostatic strain in MOCVD and MBE grown GaN films and attributed it to point defects. Based on high resolution x-ray diffraction studies of GaN films grown over $Ga_{1-x}N_x$ buffer layer on sapphire and subsequent analysis of strain data, Harutyunyan *et al.* [42] have also inferred that both hydrostatic and biaxial strains are present in these films.

We have recently shown that hydrostatic as well as biaxial strains are concurrently present in undoped epitaxial GaN films grown on *c*-sapphire by reactive sputtering at different partial pressures of $N_2$ (10% - 100%) in a growth atmosphere consisting of Ar-$N_2$ mixture; and both of these are strongly influenced by the microstructure and defects present in the films [18]. The films grown at high $N_2$ percentages (~100%) displayed large hydrostatic strain and compressive biaxial stress, which were ascribed to the presence of nitrogen interstitials. With the reduction of $N_2$ percentage in growth atmosphere, the hydrostatic strain in the films decreased substantially, along with the reversal of compressive biaxial residual stress to tensile, which was primarily ascribed to growth-related effects during crystallite coalescence. As mentioned above, in the case of epitaxial Si-doped GaN films, investigations of the dependence of residual stress on microstructure and defects and in particular, the role of intrinsic growth-related effects are quite limited. Furthermore, the study of residual stress in sputtered Si-doped GaN films and its dependence on growth atmosphere becomes more important, because the stress in sputtered films is known to be strongly influenced by the pressure and composition of growth atmosphere [43, 44]. The present study thus attempts to investigate the nature and dependence of the residual stress in sputtered epitaxial Si-doped GaN films on the partial pressure of $N_2$ in growth atmosphere. Our results reveal the concurrent presence of hydrostatic and compressive biaxial strains in the films grown at high $N_2$ percentages (>50%), which appear to be unaffected by Si incorporation. On the other hand, at lower $N_2$ percentages, the films display a dominant intrinsic tensile stress generated during coalescence, which is influenced by the microstructure and morphology of films, as well as, possible effects due to the incorporation of Si, Ar and excess nitrogen, depending upon the growth conditions.

## 2. Experimental details

Si-doped epitaxial GaN films were grown by reactive rf magnetron sputtering of GaAs and Si, as described earlier [21]. The Si area coverage of the erosion track of GaAs was fixed at ~5%, based on earlier results, which have indicated the deterioration of the structural quality of films grown with larger area coverages [21]. All the films were grown at 700 °C in Ar-$N_2$ mixture (total pressure, 0.86 Pa) over approximately 50 nm thick GaN buffer layers on *c*-sapphire substrate, which were grown in 100% $N_2$ at 300 °C. The Si-doped GaN films were grown at different nitrogen partial pressures in the range of 0.086 Pa to 0.86 Pa, by changing the corresponding $N_2$ percentage in growth atmosphere from 10% to 100%. Accordingly, the $N_2$ percentage during growth will be used to label the films grown at various partial pressures of $N_2$ in this work. The RF power was adjusted to maintain the growth rate of films at $(1.0 \pm 0.1)$ μm hr$^{-1}$, as the Ar percentage in growth atmosphere was increased. The film thickness was maintained in the range of 850 - 900 nm, implying nearly constant Ga deposition flux. With the reduction of $N_2$ in growth atmosphere from 100% to 10%, the ratio of nitrogen impingement flux to Ga deposition flux decreased from ~$2 \times 10^3$ to ~$2 \times 10^2$.

The Si content in doped GaN films was investigated by energy dispersive x-ray spectroscopy (EDX) using JEOL JSM-7600F Field Emission Gun-Scanning Electron Microscope. High resolution x-ray diffraction (HRXRD) measurements were carried out with a Cu rotating anode (9 kW), RIGAKU-Smart Lab-X-Ray diffractometer. The $a$-parameter of the films was obtained from $\omega$-$2\theta$ scans of an asymmetric reflection, which was achieved by tilting the film about $\chi$ axis, followed by the optimization of $\phi$, $\omega$, and $\chi$. The $a$-parameter was also confirmed by in-plane measurements. The $c$-parameter was obtained from $\omega$-$2\theta$ scans of symmetric (0002), (0004), and (0006) reflections. The $\omega$ and $\omega$-$2\theta$ scans of symmetric reflections were respectively used to obtain the mosaic tilt and micro-strain in the films. The mosaic twist was obtained in $\phi$-rocking curve geometry by in-plane measurement of $(11\bar{2}0)$ reflection, using parallel slit collimator (PSC_1°) and analyzer (PSA_1°). Atomic force microscopy (AFM) was caried out in contact mode with Nanoscope IV Multimode scanning probe to study the surface morphology over an area of 2 μm × 2 μm.

## 3. Results and Discussion

The composition of Si-doped GaN films was investigated by EDX and the corresponding data (along with that of a commercially procured, MOCVD grown GaN film) is presented in Fig. S1 and Table S1 of supplementary material. It is seen that the measured N/Ga ratio in commercial GaN film is 0.70, while that for the Si-doped GaN films decreases from 0.76 to 0.69, with the reduction of $N_2$ in growth atmosphere from 100% to 10%. The EDX data also confirm the absence of arsenic and the presence of oxygen impurity (~3 at.%) in all the films. These results are in agreement with earlier x-ray photoelectron spectroscopy (XPS) and secondary ion mass spectroscopy (SIMS) studies of undoped GaN films [18, 45], which also indicated the possibility of excess nitrogen in the films grown at higher $N_2$ percentages. All the films, which were grown with 5% Si area coverage, reveal the presence of ~2 at.% Si, irrespective of $N_2$ percentage in growth atmosphere. Phi ($\phi$) scans of $(10\bar{1}1)$ reflection of GaN were carried out to assess the epitaxial character and quality of Si-doped GaN films obtained at various $N_2$ percentages in growth atmosphere and are shown as Fig. S2 of supplementary material. All the films exhibit the six $(10\bar{1}1)$ peaks at intervals of 60°. The $\phi$-scan of $(10\bar{1}4)$ peaks of sapphire, which is also shown in the figure, reveals that these peaks are rotated by 30° with respect to the GaN peaks, thus conforming to the in-plane epitaxial relationship of GaN$[11\bar{2}0] \| \alpha$-$Al_2O_3[10\bar{1}0]$ [46]. These results indicate the alignment of GaN lattice with the oxygen sub-lattice of sapphire and confirm the epitaxial character and mosaic structure of the films.

Figure 1(a) shows $\omega$-$2\theta$ high-resolution scans for symmetric (0002), (0004) and (0006) reflections of GaN, from which the average values of $c$-parameter for all the films were obtained, with an uncertainty of $\pm 0.001$ Å. For the determination of $a$-parameter of the films, the $\omega$-$2\theta$ high-resolution $(10\bar{1}1)$ asymmetric scans and the in-plane measurements of $(11\bar{2}0)$ and $(20\bar{2}0)$ reflections were carried out. Figure 1(b) shows the $\omega$-$2\theta$ asymmetric scans of $(10\bar{1}1)$ reflection for all the films, while the corresponding in-plane data for $(11\bar{2}0)$ and $(20\bar{2}0)$ reflections are shown in supplementary material as Fig. S3. These measurements were used to obtain the average values of $a$-parameter of the films, with an uncertainty of $\pm 0.002$ Å. The reproducibility of measurements was confirmed by analyzing multiple films grown at each $N_2$ percentage. Fig. S4 shows the values of the lattice parameters, $a$ and $c$ (average with error bars), which are plotted against Ar and $N_2$ percentages in growth atmosphere.

The microstructural parameters of the films, namely, the out-of-plane micro-strain, mosaic tilt and twist were obtained by HRXRD measurements and these results are shown for typical cases, in Figs. 2 – 4 and are summarized in Fig. 5. Figure 2(a) shows the $\omega$-$2\theta$ scans of symmetric (0002), (0004) and (0006) reflections and the Williamson-Hall plots of the corresponding reciprocal space broadening, $\Delta q_z$ versus the magnitude of reciprocal lattice vector (q) are shown in Fig. 2(b), from which the out-of-plane micro-strain was obtained. A relatively large full width at half maximum (FWHM) of $\omega$-$2\theta$ peaks is seen in the film grown at 100% $N_2$, which decreases initially with the reduction of $N_2$ percentage, as is typically shown for 50% $N_2$. However, the FWHM's of $\omega$-$2\theta$ peaks are found to increase slightly at lower $N_2$ percentages. Figure 3(a) shows the $\omega$ scans of symmetric (0002), (0004) and (0006) reflections and the Williamson-Hall plots of the corresponding reciprocal space broadening, $\Delta q_x$ versus q are shown in Fig. 3(b), from which the mosaic tilt was obtained. In this case also, the film grown at 100% $N_2$ displays a relatively large FWHM of $\omega$ scans, which decreases substantially for films grown in the range of 50% - 75% $N_2$, but again displays a slight increase, as is shown typically for 30% and 10% N$_2$. Figure 4 shows the in-plane $\phi$-rocking curves of (11$\bar{2}$0) reflection, from which, the mosaic twist of the films was obtained. The $\phi$-rocking curve of film grown at 100% N$_2$ displays a large FWHM, which decreases slightly to nearly the same values for films grown in the range of 30% - 75% N$_2$. However, below 30% N$_2$, the FWHM decreases substantially (shown typically for 10% N$_2$), indicating a significantly reduced mosaic twist.

The mosaic tilt and twist values obtained from above were used to calculate the densities of screw and edge dislocations [47], which along with the micro-strain are plotted in Fig. 5, against Ar and N$_2$ percentages in growth atmosphere. It may be mentioned here that in hexagonal systems, including GaN, the density of edge dislocations dominates over the density of screw dislocations [48, 49], which was also observed in the case of undoped sputtered GaN epitaxial films [18]. Figure 5 shows high densities of edge ($\sim 3 \times 10^{12}$ cm$^{-2}$) and screw ($\sim 5 \times 10^{11}$ cm$^{-2}$) dislocations in the film grown at 100% N$_2$, both of which decrease slightly as N$_2$ is reduced to the range of 30% - 50%. Interestingly, below 30% N$_2$, the density of edge dislocations continues to decrease, but that of screw dislocations tends to increase, thus both attaining comparable values in the range of (3 - 5) $\times 10^{11}$ cm$^{-2}$, for the film grown at 10% N$_2$. Figure 5 also shows that a large micro-strain ($\sim 1.5 \times 10^{-2}$) is present in the film grown at 100% N$_2$. As N$_2$ is reduced to $\sim$50%, the micro-strain decreases substantially ($\sim 2 \times 10^{-3}$), but again increases marginally with reduction of N$_2$ below 30%. The presence of substantial micro-strain is attributed to the high density of screw dislocations in these films [50]. However, as the sputtering of GaAs was carried out in Ar-N$_2$ atmosphere of a widely varying composition (10% - 100% N$_2$), there is also a possibility of the incorporation of excess nitrogen and/or argon in the films, depending on the N$_2$/Ar ratio, which may also contribute to the micro-strain [51, 52], as will be discussed below.

The morphology of Si-doped GaN films grown at different N$_2$ percentages in growth atmosphere was also investigated and the results are shown for typical cases in Fig. S5 of supplementary material, along with the corresponding root mean square (rms) roughness of the films. The morphology of films grown in the range of 30% - 100% N$_2$ are nearly similar, displaying uniform and well-rounded lateral surface features, along with rms roughness of 3-5 nm. For the films grown at 20% - 30% N$_2$, the lateral features become asymmetric and irregular, the surface morphology becomes slightly voided and rms roughness increases substantially. Interestingly, in the films grown at lower than 20% N$_2$, well-rounded, symmetric lateral surface features are again seen, but with a substantially voided surface morphology, which results in the high values of their rms roughness (15-20 nm). Since the growth rate of the films was nearly constant, the changes in surface morphology with reduction of $N_2$ percentage are ascribed to the increasing effect of Ar bombardment [53].

The values of the measured in-plane strain ($\varepsilon_1 = (a - a_0)/a_0$) and out-of-plane strain ($\varepsilon_2 = (c - c_0)/c_0$) were obtained from the data in Fig. S4 and are plotted in Fig. 6, against Ar and $N_2$ percentages in growth atmosphere, where, $a_0$ (3.189 Å) and $c_0$ (5.185 Å) represent the unstrained lattice parameters. The extrinsic in-plane (biaxial) strain ($\varepsilon_{th} \simeq -1.28 \times 10^{-3}$), due to the difference in coefficients of thermal expansion of GaN and $c$-sapphire ($5.59 \times 10^{-6}$ K$^{-1}$ and $7.5 \times 10^{-6}$ K$^{-1}$, respectively) corresponding to $\Delta T = 670$ K (the difference between growth and room temperatures) is compressive in the present case, and is also shown in Fig. 6 by a horizontal line. The film grown at 100% $N_2$ shows a relatively large out-of-plane tensile strain (~$5 \times 10^{-3}$), which becomes slightly compressive in those grown in the range of 20% - 50% $N_2$, but regains mildly tensile character at lower $N_2$ percentages. On the other hand, the in-plane strain in the film grown at 100% $N_2$ is compressive ($-2.2 \times 10^{-3}$), which becomes tensile as $N_2$ is reduced below 75%. It attains the highest value of ~$2.5 \times 10^{-3}$ at 30% $N_2$, but decreases substantially at lower $N_2$ percentages.

It is noted from Fig. 6 that the measured in-plane and out-of-plane strains do not display solely hydrostatic ($\varepsilon_1 = \varepsilon_2$) or biaxial ($\varepsilon_2/\varepsilon_1 \simeq -0.53$ for GaN [54, 55]) character in any of the films, indicating that both these contributions to strain are concurrently present. It is known that point defects like, $V_{Ga}$, $V_N$, $Ga_i$, $N_i$, $N_{Ga}$, $Ga_{N,}$ [56] and O interstitials [57] are present in epitaxial GaN films. As mentioned above, there is also a possibility of the substantial incorporation of excess nitrogen and/or argon in the films, depending on the $N_2$/Ar ratio during growth. The hydrostatic strain may in general, originate from the above mentioned defects and impurities, causing expansive or compressive strain in wurtzite GaN. On the other hand, the in-plane biaxial strain can be extrinsic, due to the substantial difference between the coefficients of thermal expansion of GaN and sapphire or can also be intrinsic, due growth-related effects, including the in-plane tensile stress generated during coalescence. Thus, considering the concurrent presence of hydrostatic and biaxial strains, the measured strain data of these films and its dependences on $N_2$/Ar ratio during growth have been quantitatively analyzed. In such a situation, the measured in-plane and out-of-plane components of strain ($\varepsilon_1$ and $\varepsilon_2$), are given by [40, 42]

$$\varepsilon_1 = \varepsilon_{1b} + \varepsilon_h \qquad (1)$$

$$\varepsilon_2 = \varepsilon_{2b} + \varepsilon_h \qquad (2)$$

where, $\varepsilon_{1b}$ and $\varepsilon_{2b}$ are the biaxial strain components along $a$ and $c$ directions, respectively, while, $\varepsilon_h$ is the hydrostatic strain, given by

$$\varepsilon_h = \frac{1-\nu}{1+\nu}\left(\varepsilon_2 + \frac{2\nu}{1-\nu}\varepsilon_1\right) \qquad (3)$$

and $\nu$ {$= c_{13}/(c_{13} + c_{33})$}is the Poisson ratio. For GaN, $c_{13} = 106$ GPa and $c_{33} = 398$ GPa are the values of elastic constants and $\nu$ is ~0.21 [42, 54]. Equations (1)-(3) were used to obtain the hydrostatic and biaxial strain contributions in all the films. The effect of thermal strain contribution ($\varepsilon_{th}$) was eliminated from the biaxial strain component ($\varepsilon_{1b}$), which is elastically related to the biaxial strain component $\varepsilon_{2b}$ (= -($2c_{13}/c_{33})\varepsilon_{1b}$) [55], by subtracting $\varepsilon_{th}$ from $\varepsilon_{1b}$. Thus, $\varepsilon_{1b}^-$ (= $\varepsilon_{1b}$ - $\varepsilon_{th}$), contains only the intrinsic contribution to in-plane biaxial strain and is hence, not commensurate with $\varepsilon_{2b}$. The intrinsic biaxial stress which arises entirely due to growth-related effects can thus be related to $\varepsilon_{1b}^-$ as

$$\sigma_f = M_f \varepsilon_{1b}^- \qquad (4)$$

where, $M_f$ is the biaxial elastic modulus, which is given by { $c_{11} + c_{12}$ -2($c_{13}^2/c_{33}$)}and obtained from the elastic constants $c_{11}$ (= 390 GPa) and $c_{12}$ (= 145 GPa) of GaN [42, 54].

Figure 7 shows the variations of $\varepsilon_h$, $\varepsilon_{1b}^-$, $\varepsilon_{2b}$ and $\sigma_f$ with Ar and $N_2$ percentages in growth atmosphere and their behaviour is summarized below. The plot of $\varepsilon_h$ in Fig. 7(a) shows a large hydrostatic strain (~$2.5 \times 10^{-3}$) in the film grown at 100% $N_2$. As $N_2$ is reduced to 50%, $\varepsilon_h$ decreases below $10^{-3}$ and increases slightly below 30% $N_2$. The corresponding plots of $\varepsilon_{1b}^-$ and $\varepsilon_{2b}$ are shown in Fig. 7(b). A large value of $\varepsilon_{2b}$ (~$2.5 \times 10^{-3}$) is seen in the film grown at 100 % $N_2$, implying a tensile out-of-plane strain. Interestingly, with reduction of $N_2$ to ~30%, $\varepsilon_{2b}$ changes its character from tensile to compressive (-$10^{-3}$), but again becomes slightly tensile (~$3.5 \times 10^{-4}$) in the film grown at 10% $N_2$. The intrinsic biaxial stress ($\sigma_f$) was obtained from the growth-related, intrinsic in-plane strain ($\varepsilon_{1b}^-$) in Fig. 7(b), and is plotted in Fig. 7(c), which shows that the film grown at 100% $N_2$ possesses a relatively large compressive intrinsic stress $\sigma_f \approx$ - 1.6 GPa ($\varepsilon_{1b}^- \approx$ - $3.4 \times 10^{-3}$). However, the intrinsic compressive stress $\sigma_f$ reverses its character to tensile below 75% $N_2$ and remains so in the films grown in the range of 30% - 50% $N_2$, with $\sigma_f$ being in the range of 0.9 - 1.7 GPa ($\varepsilon_{1b}^-$ of (2 - 3) $\times 10^{-3}$). However, $\sigma_f$ decreases to $\approx 0.3$ GPa, as $N_2$ in growth atmosphere is further reduced to ~10%.

It is evident from above that the Si-doped GaN films grown at high $N_2$ percentages (75% - 100 %) display large hydrostatic strain (>$10^{-3}$), which relaxes substantially with the reduction of $N_2$ percentage and tends to increase again in the films grown at 10% - 20% $N_2$. Interestingly, this behaviour is quite similar to that of the micro-strain, as shown in Fig. 5. The hydrostatic strain in GaN films, which is characterized by the concomitant increase of $a$ and $c$ has been ascribed to the presence of self-interstitials [58], as was also reported recently in undoped GaN films grown by sputtering [18]. As the EDX measurements have indicated, the films grown at $N_2$ percentages $\gtrsim$75% are likely to contain excess/interstitial nitrogen, which decreases with the reduction of $N_2$ percentage in growth atmosphere. Thus, in the film grown at 100% $N_2$ which has the highest N/Ga ratio, the large hydrostatic strain and micro-strain are attributed to excess/interstitial nitrogen; and their subsequent decrease at lower $N_2$ percentages to the reduction of N/Ga ratio. Figure 7(a) also shows a slight increase of hydrostatic strain in the films grown at 10% - 20% $N_2$, as is also seen in the case of micro-strain in Fig. 5, which is attributed to the incorporation of Ar in these films, as these are grown at high Ar percentages (80% - 90%). There is no evidence of compressive hydrostatic effect due to the incorporation of Si in these films, which is in tune with the theoretical predictions mentioned above; and in any case, it is likely to be masked by the dominant and expansive hydrostatic strain due to the presence of interstitial nitrogen and/or argon.

Figure 7(c) shows that that an intrinsic biaxial stress is present in Si-doped GaN films, which in most of the cases, is tensile in character, except for the films grown at high $N_2$ percentages (75% - 100%). This is in agreement with the studies on MOCVD grown Si-doped GaN films described above, in which a tensile biaxial stress has been usually reported. This behaviour is also similar to that observed in our earlier work [18] on undoped GaN films grown by sputtering, which was attributed to the growth-related tensile stress generated during coalescence. The growth-related intrinsic tensile stress in polycrystalline thin films is known to develop during crystallite coalescence, due to the attraction between crystallites [39]. Nix and Clemens [39] have proposed a quantitative model under low adatom mobility condition, and have asserted that the tensile stress developed during crystallite coalescence is maintained during film growth. It has also been reported [59, 60] that edge type threading dislocations of high density ($\sim 10^{10}$ cm$^{-2}$) are formed at the coalescence stage during the mosaic growth of epitaxial GaN films, which subsequently lead to low-angle grain boundaries that are rotated about the *c*-axis. Potin *et al*. [60] have also observed high-angle grain boundaries arising from the local arrangements of threading dislocations, which extend throughout the thickness of epitaxial GaN films. It appears that the high density of dislocations in sputtered Si-doped GaN films which originate during the initial stages of growth, lead to the formation of low and high angle grain boundaries, manifesting in the significant twist and tilt of mosaic blocks. It is thus inferred that the tensile stress seen in most of the sputtered Si-doped GaN films originates during crystallite coalescence, as was also suggested by Romano *et al.* [29] for MOCVD grown films. Although, the incorporation of Si on the compressive side of edge dislocations has also been suggested to cause tensile in-plane stress [38], this effect appears to be marginal in the present case, owing to the similarity of the strain behaviour of Si-doped GaN films with that of undoped GaN films, which was reported earlier [18].

Interesting features are however seen in the behaviour of intrinsic biaxial stress at the higher and lower ends of $N_2$ percentage. Firstly, in the film grown at 100% $N_2$, the intrinsic stress is compressive with a substantially large value (-1.6 GPa), but reverses to tensile character in the films grown below 75% $N_2$. Secondly, at the lower end of $N_2$ percentage, the tensile intrinsic stress, tends to decrease substantially. Polycrystalline metal films deposited by sputtering are known to display a reversal of in-plane stress from tensile to compressive stress at low sputtering pressure [43, 44, 61, 62] and high negative substrate bias [61, 63], which has been ascribed to the incorporation of inert atoms [43, 44] and oxygen [64]. In the present work, all the Si-doped GaN films were grown at a fixed sputtering pressure, but as mentioned above, the N/Ga ratio is substantially larger in the films grown at higher $N_2$ percentages, while the Si and O contents of the films remain practically unchanged. It is thus likely that the source of the compressive intrinsic stress in the films grown at high $N_2$ percentages is the same as that invoked for explaining the large hydrostatic strain. It is thus inferred that incorporation of excess nitrogen into grain boundaries is the source of compressive stress in the films grown at high $N_2$ percentages [39, 65], while another possible cause could be the attraction of nitrogen interstitials towards tensile side of edge dislocations [66]. Thus, the large compressive stress in Si-doped GaN films grown at higher $N_2$ percentages (~100%), is primarily attributed to the presence of excess/interstitial nitrogen, which is in tune with our earlier studies on undoped GaN films [18]. With the decrease in N/Ga ratio in the films grown at 30% - 50% $N_2$, the effects due to excess nitrogen are suppressed, resulting in the reversal of intrinsic stress, which is now determined primarily by the coalescence related intrinsic tensile stress. Finally, in the films grown below 30% $N_2$, a reduction of tensile stress is seen, which may be explained as the relaxation of coalescence related tensile stress by the incorporation of Ar in the films, which, as mentioned above, cannot be ruled out during the growth at high Ar percentages [44]. Thus, the incorporation of Ar in the films grown at low $N_2$ percentages consistently explains the increase in hydrostatic strain and micro-strain, together with the decrease of intrinsic tensile stress. It may also be mentioned that these films have a substantially voided morphology and large surface roughness, which may also possibly contribute to the reduction of intrinsic tensile stress, as observed in zone 1 type sputtered metal films [67, 68].

## 4. Conclusion

Hetero-epitaxial Si-doped GaN films have been grown on sapphire by reactive co-sputtering of GaAs and Si in Ar-$N_2$ growth atmosphere. The residual stress in the films grown at different $N_2$ percentages in the range of 10% - 100% have been analyzed and correlated with microstructural parameters. EDX results show that all the films contain ~2 at.% Si and ~3 at.% oxygen impurity, while the N/Ga ratio decreases with reduction of $N_2$ percentage in growth atmosphere. The in-plane and out-of-plane strain components obtained from independent measurements of $a$ and $c$ parameters are strongly dependent on $N_2$ percentage and reveal the concurrent presence of hydrostatic and biaxial strain contributions. The films grown at 30%-100% $N_2$ show the dominant presence of edge dislocations (density ~$10^{12}$ cm$^{-2}$), which decrease to ~$5 \times 10^{11}$ cm$^{-2}$ in the films grown at ~10% $N_2$. However, below 30% $N_2$, the density of screw dislocations increases and approaches values comparable to those of edge dislocations. The film grown at 100% $N_2$ also displays large micro-strain (~$1.5 \times 10^{-2}$) and hydrostatic strain (~$2.5 \times 10^{-3}$) due to the incorporation of nitrogen at interstitial locations, as well as growth-related compressive biaxial stress (-1.6 GPa), which is attributed to nitrogen incorporation into the grain boundaries and possibly tensile side of edge dislocations. The micro-strain and hydrostatic strain decrease substantially with the decrease of N/Ga ratio in the films grown below 75% $N_2$, due the decrease of excess nitrogen, resulting in the prevalence of the intrinsic tensile stress, which is generated during coalescence. In the case of the films grown below 30% $N_2$, the slight increase of micro-strain and hydrostatic strain as well as the reduction of intrinsic tensile stress are attributed to the incorporation of Ar, with the latter being also influenced by possibly the voided morphology of films. An interesting observation from the present work on sputtered Si-doped GaN films is that the high doping concentration of Si does not appear to have any hydrostatic influence on the strain behaviour, which is in tune with theoretical predictions, as well as most of the earlier reported measurements on MOCVD grown films. Our results however show that in any case, the effect of Si incorporation is likely to be masked by growth-related stresses due to incorporation of nitrogen and/or argon in the films, and more importantly, by the intrinsic tensile stress generated during coalescence, which is dominant in most of the films.

## CRediT authorship contribution statement

**Mohammad Monish:** Conceptualization, Methodology, Data curation, Investigation, Visualization, Writing - original draft. **S. S. Major:** Conceptualization, Methodology, Supervision, Writing - review & editing.

**Declaration of competing interest**

The authors declare that they have no known competing financial interests or personal relationships that could have appeared to influence the work reported in this paper.

**Acknowledgements**

The HRXRD and AFM Central Facilities as well as Sophisticated Analytical Instruments Facility (SAIF) of IIT Bombay are greatly acknowledged for respective measurements.

**Data Availability**

The data that support the findings of this study are available from the corresponding author upon reasonable request.

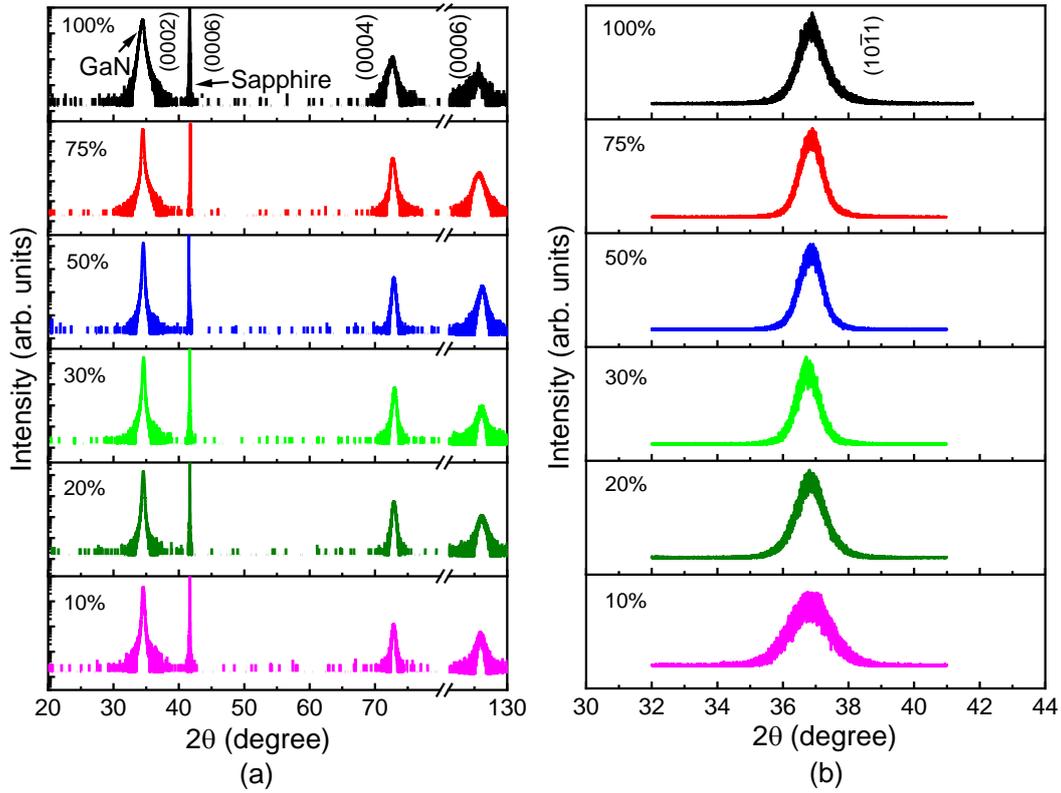

**Fig. 1.** The out-of-plane $\omega$-$2\theta$ scans of (a) symmetric (0002), (0004) and (0006) reflections (plotted in log scale) and (b) asymmetric ($10\bar{1}1$) reflection for Si-doped GaN films grown on *c*-sapphire at different $N_2$ percentages in growth atmosphere (as indicated).

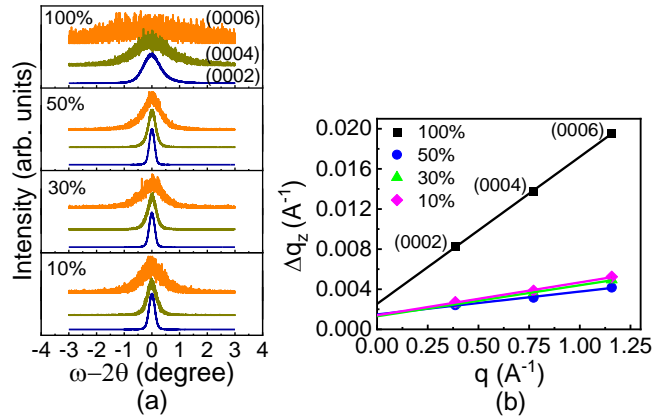

**Fig. 2.** (a) The out-of-plane $\omega$-$2\theta$ scans of (0002), (0004), and (0006) reflections and (b) the plot of $\Delta q_z$ (along $\omega$-$2\theta$ axis) *vs.* q, for typical Si-doped GaN films grown on *c*-sapphire at different $N_2$ percentages in growth atmosphere (as indicated).

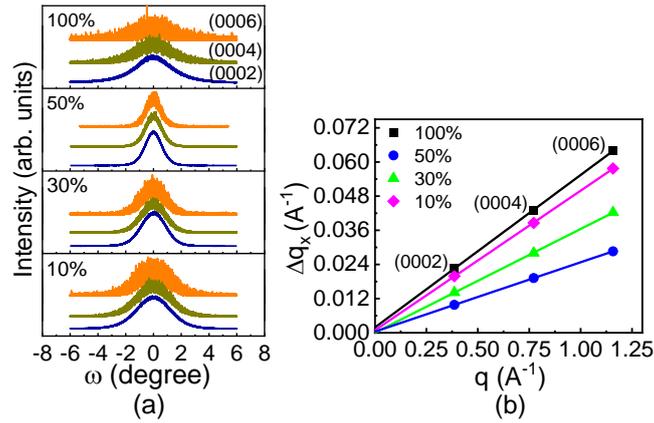

**Fig. 3.** (a) The $\omega$ scans of (0002), (0004), and (0006) reflections and (b) the plot of $\Delta q_x$ (along $\omega$ axis) *vs.* q for typical Si-doped GaN films grown on *c*-sapphire at different N₂ percentages in growth atmosphere (as indicated).

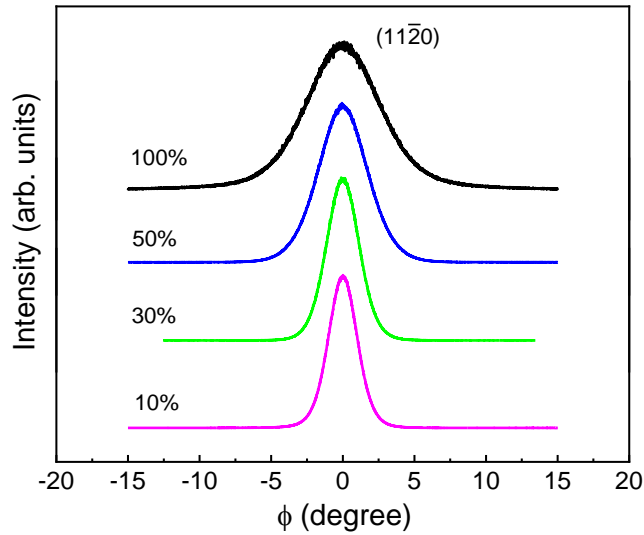

**Fig. 4.** In-plane $\phi$-rocking scans of (11$\bar{2}$0) reflection for typical Si-doped GaN films grown at different N₂ percentages in growth atmosphere (as indicated).

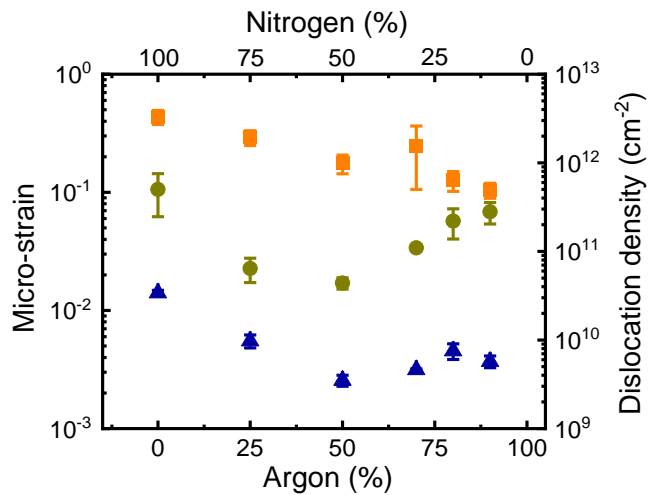

**Fig. 5.** The plots of micro-strain ( ▲ ), density of edge dislocations ( ■ ) and density of screw dislocations ( ● ) in Si-doped GaN films grown at different Ar /N₂ percentages in growth atmosphere.

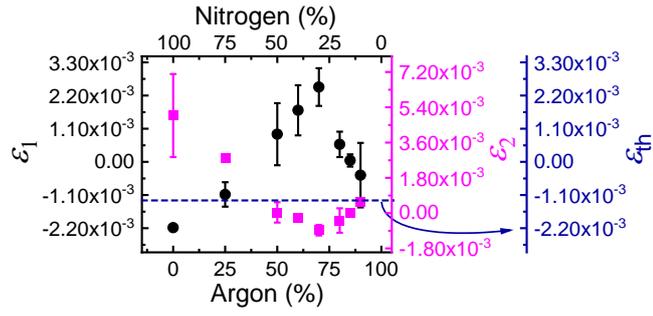

**Fig. 6.** The plots of measured in-plane strain, $\varepsilon_1$ (●) and out-of-plane strain, $\varepsilon_2$ (■) in Si-doped GaN films grown at different Ar/N$_2$ percentages in growth atmosphere. The dashed horizontal line represents the in-plane thermal strain ($\varepsilon_{th}$) due to the difference between the coefficients of thermal expansion of GaN and $c$-sapphire.

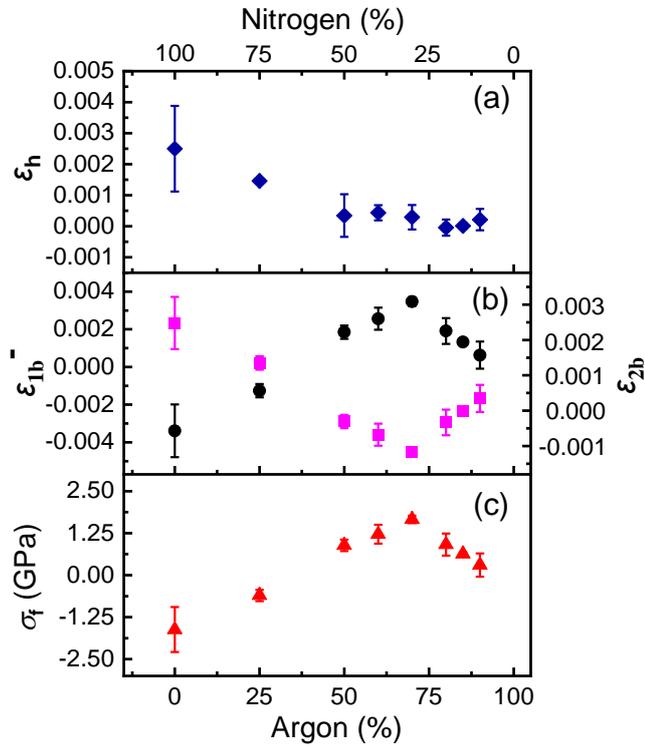

**Fig. 7.** The calculated (a) hydrostatic strain, $\varepsilon_h$ (◆), (b) biaxial strains, $\varepsilon_{1b}^-$ (●) and $\varepsilon_{2b}$ (■), and (c) the growth-related intrinsic biaxial stress, $\sigma_f$ (▲) in Si-doped GaN films grown at different Ar/N$_2$ percentages in growth atmosphere.

**Supplementary material**

**Study of residual stress in reactively sputtered epitaxial Si-doped GaN films**

Mohammad Monish[*] and S. S. Major[†]

*Department of Physics, Indian Institute of Technology Bombay, Mumbai 400076, India*

## 1. Energy dispersive X-ray spectroscopy

Energy dispersive X-ray spectroscopy (EDX) was carried out to study the composition of Si-doped GaN films, and in particular, evaluate their Si content and confirm the absence of arsenic. The results are shown in Fig. S1 and the data obtained from EDX analysis are presented in Table S1, which show that all the films contain ~2 at.% Si. It is found that the Si content of the films remains in the range of 2-3 at.%, irrespective of the increase of Si area coverage, which results in deterioration of the structural quality of films.

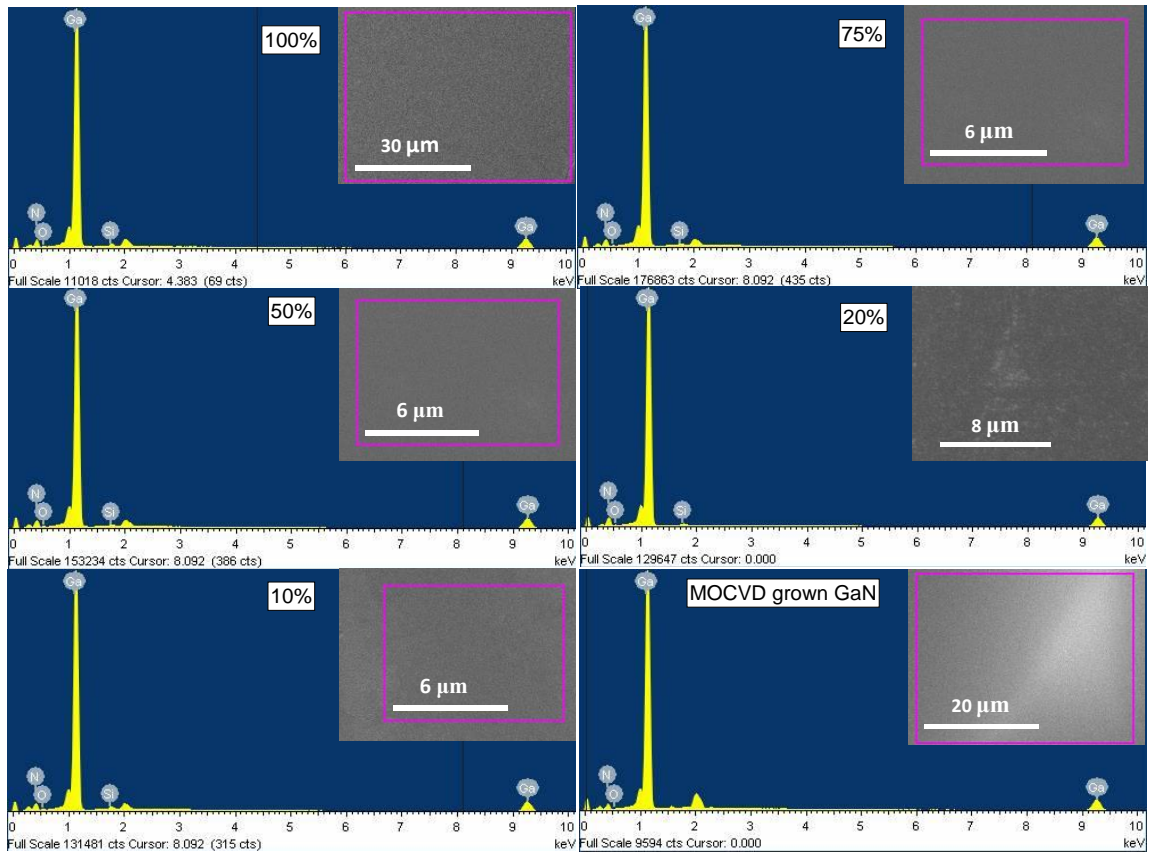

**Fig. S1.** EDX results of Si-doped GaN films grown on *c*-sapphire at different $N_2$ percentages in growth atmosphere (as indicated), which also include the data for a commercially procured, MOCVD grown undoped GaN film.

---

[*] monish.iitb@gmail.com

[†] Author to whom any correspondence should be addressed: syed@iitb.ac.in

**Table S1.** Summary of EDX analysis of Si-doped GaN films grown at different $N_2$ percentages in growth atmosphere along with the data for commercially procured MOCVD grown undoped GaN.

| $N_2$ (%) | Ga at. (%) | N at. (%) | Si at. (%) | O at. (%) | N/Ga |
|---|---|---|---|---|---|
| 100 | 54.33 | 41.03 | 1.92 | 2.72 | 0.76 |
| 75 | 54.78 | 41.17 | 1.47 | 2.58 | 0.75 |
| 50 | 54.50 | 40.97 | 1.65 | 2.88 | 0.75 |
| 20 | 56.04 | 39.45 | 1.54 | 2.97 | 0.70 |
| 10 | 55.73 | 38.88 | 1.85 | 3.54 | 0.69 |
| MOCVD grown GaN | 58.36 | 40.87 | - | 0.77 | 0.70 |

## 2. Phi (ϕ) scans

The phi (ϕ) scans of ($10\bar{1}1$) reflections for Si doped GaN films grown at different $N_2$ percentages, along with the ϕ-scan of ($10\bar{1}4$) reflections of sapphire are shown in Fig. S2, which show six dominant peaks at intervals of 60° with each other, corresponding to ($10\bar{1}1$) reflections. The ϕ-scan of ($10\bar{1}4$) reflections of sapphire are rotated by 30°, which imply the in-plane epitaxial relationship of GaN[$11\bar{2}0$]∥α-Al$_2$O$_3$[$10\bar{1}0$].

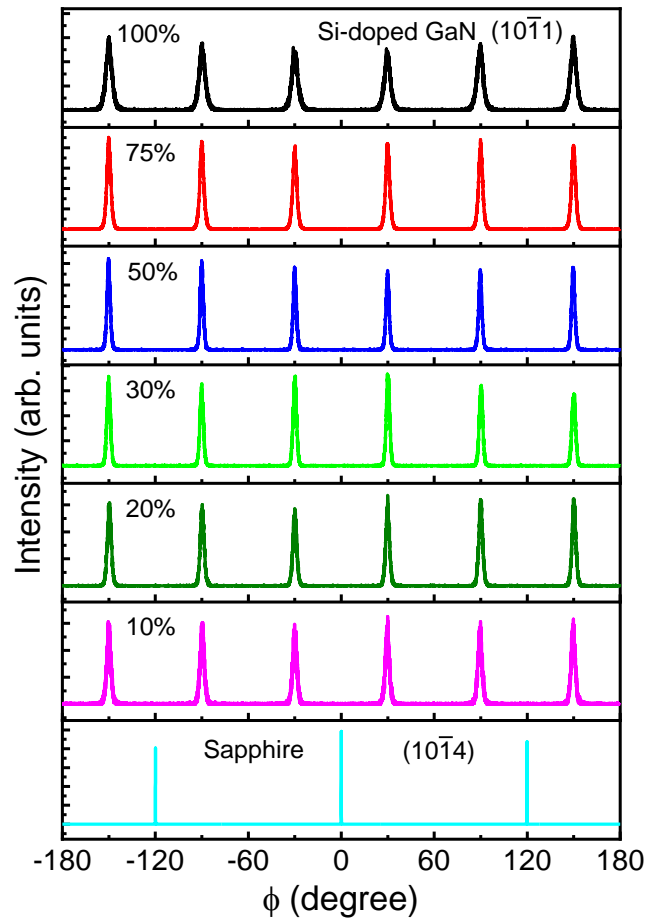

**Fig. S2.** ϕ-scans of ($10\bar{1}1$) reflections of Si-doped GaN films grown on *c*-sapphire at different $N_2$ percentages in growth atmosphere (as indicated), along with the ϕ-scan of ($10\bar{1}4$) reflections of *c*-sapphire.

### 3. In-plane x-ray diffraction

The in-plane measurements of $(11\bar{2}0)$ and $(20\bar{2}0)$ reflections were used to obtain the $a$-parameter of Si-doped GaN films grown at different $N_2$ percentages in sputtering atmosphere. These results are shown in Fig. S3.

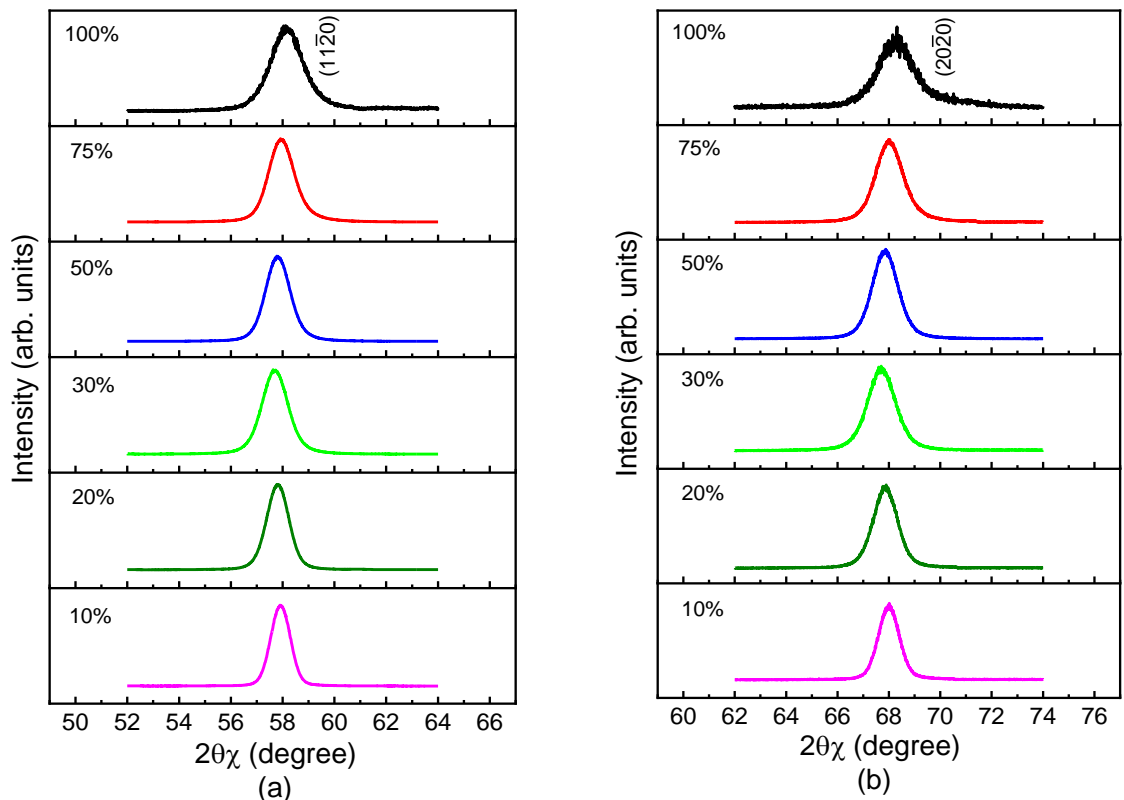

**Fig. S3.** In-plane measurement of (a) $(11\bar{2}0)$ and (b) $(20\bar{2}0)$ reflections for Si-doped GaN films grown on $c$-sapphire at different $N_2$ percentages in growth atmosphere (as indicated).

### 4. Lattice parameters

The $\omega$-$2\theta$ high resolution scans of symmetric and asymmetric reflections as well as in-plane measurements were carried out to independently obtain the $c$ and $a$ parameters of epitaxial Si-doped GaN films grown at different $N_2$ percentages. The average values of $c$ and $a$ parameters along with the error bars are plotted in Fig. S4 against Ar and $N_2$ percentages in growth. The film grown at 100% $N_2$ shows a significantly larger $c$-parameter (5.211 ± 0.01) Å and a marginally smaller $a$-parameter (3.182 ± 0.0002) Å, compared to the corresponding standard values for bulk GaN ($c$ = 5.185 Å, $a$ = 3.189 Å). With decrease of $N_2$ percentage in growth atmosphere, the $c$-parameter decreases monotonically and becomes (5.181 ± 0.001) Å for the film grown at 30% $N_2$, which is slightly smaller than the standard value. However, with further decrease of $N_2$ percentage, the $c$-parameter increases again, becoming slightly larger than the standard value. On the other hand, the $a$-parameter increases initially with decrease of $N_2$ percentage and becomes substantially larger than the standard value for the film grown at 30% N₂. However, with further decrease of N₂ percentage (<30%), the *a*-parameter decreases substantially, displaying values comparable to the standard value.

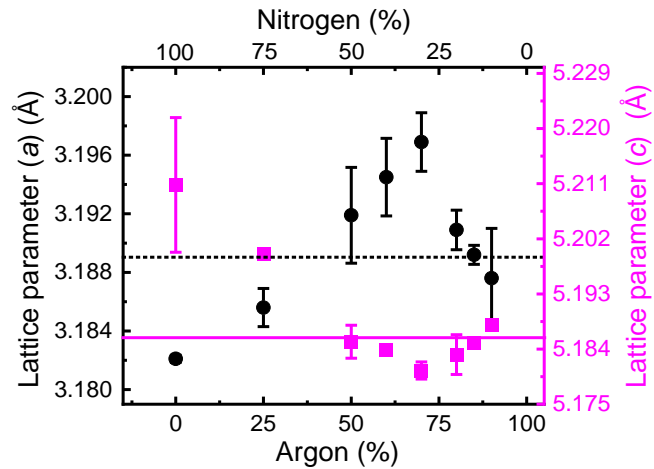

**Fig. S4.** *a*-parameter (●) and *c*-parameter (■) of Si-doped GaN films plotted against Ar/N₂ percentages in growth atmosphere. The horizontal lines (……) and (———) represent the standard values of *a* and *c* parameters, respectively.

## 5. Surface morphology

The surface morphology of Si-doped GaN films grown at different N₂ percentage was investigated by AFM in contact mode and the results are shown in Fig. S5 for typical cases, along with corresponding root mean square (rms) surface roughness data plotted against Ar/N₂ percentages in sputtering atmosphere.

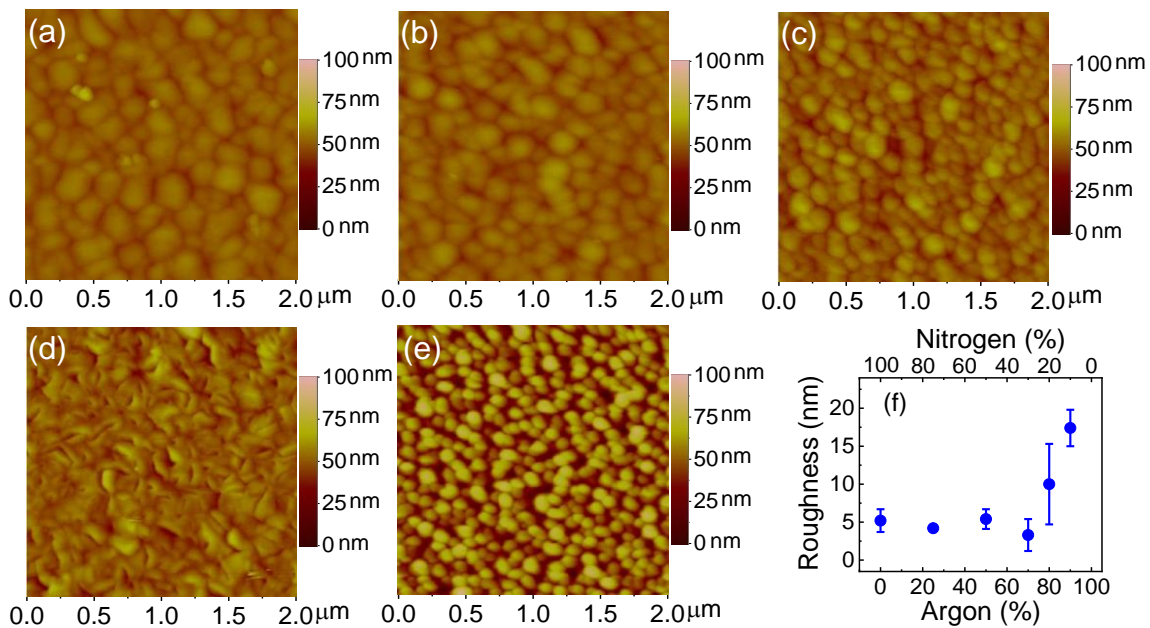

**Fig. S5.** Top view AFM images of typical Si-doped GaN films grown at (a) 100%, (b) 50%, (c) 30%, (d) 20%, and (e) 10% N₂ in growth atmosphere, along with (f) the rms surface roughness (●) of Si-doped GaN films plotted against Ar/N₂ percentages in growth atmosphere.